# Physical Realization of von Neumann Lattices in Rotating Dipole-blockaded Bose Gases

Szu-Cheng Cheng[1] & Shih-Da Jheng [2]

A mathematical lattice, called the von Neumann lattice[1], is a subset of coherent states and exists periodically in the phase space. It is unlike solids or Abrikosov lattices[2] that are observable in physical systems. Abrikosov lattices are vortices closely packed into a lattice with a flux quantum through a unit cell. Although Abrikosov lattices appear generally in various physical systems[3–10], vortex lattices with multiple-flux quantums through a unit cell are more stable than Abrikosov lattices in some physical regimes of the systems with non-local interactions between particles[11,12]. No theory is able to describe these vortex lattices today. Here, we develop a theory for these vortex lattices by extending von Neumann lattices to the coordinate space with a unit cell of area that is proportional to flux quantums through a unit cell. The von Neumann lattices not only show the same physical properties as the Abrikosov lattice, but also describe vortex lattices with multiple-flux quantums through a unit cell. From numerical simulations of a rapidly rotating dipole-blockaded gas, we confirm that vortex lattices showed in our simulations are the representation of von Neumann lattices in the coordinate space. We anticipate our theory to be a starting point for developing more

*[1]Dept. of Optoelectric Physics, Chinese Culture University, Taipei 11114, Taiwan, ROC; [2]Institute of Physics, National Chiao Tung University, Hsinchu 30010 ,Taiwan, ROC*

**sophisticated vortex-lattice models. For example, the effect of Landau-level mixing on vortex lattice structures[11], vortices formed inside superfluid droplets[12] and structural phase transitions of vortex matter in two-component Bose-Einstein condensates[13] will be relevant for such developments.**

Rotating gases experience a centrifugal force, called the Coriolis force, in the rotating frame. The force acting on two-dimensional (2D) rotating gases with a rotational frequency $\Omega$ is similar to the Lorentz force experienced by charged particles in a uniform magnetic field $B$, $B = 2\Omega$, which is perpendicular to a 2D plane and along the z-direction, $\hat{\mathbf{z}}$. Quantum-mechanically, a particle feels a gauge field in a magnetic field and its wave functions and energy levels can be solved by choosing a particular gauge of the magnetic field. For a Landau gauge, energy levels and wave functions of a particle in a uniform magnetic field are discrete Landau levels and Landau wave functions[14], respectively. Abrikosov[2] constructed a macroscopic wave function, expanded by periodical Landau wave functions in the lowest Landau level (LLL), to predict the existence of a vortex lattice with a flux quantum through a unit cell. This lattice is named as the Abrikosov lattice and occurs generally in type-II superconductors[3,4], superfluid helium[5,6], Bose-Einstein condensates[7,8], ultracold fermion superfluids[9], and dark matter condensates[10]. Although Abrikosov lattices appear in various physical systems, there is no generalization of these lattices to vortex lattices with multiple-flux quantums through a unit cell.

The Landau gauge was selected to describe the physical properties of Abrikosov lattices. Some people choose the symmetric gauge in studying these lattices[15]. In quantum Hall effects, the symmetric gauge was also used to describe the correlation of electrons in the LLL[16]. In the symmetric gauge, a continuous set of coherent states in the LLL can realize the cyclotron motion of a particle in a magnetic field. These states are not orthogonal each other and overcomplete[14]. The overcompleteness means that a subset being complete belongs to one of the subsets of coherent states. von Neumann[1] found that a set of periodical coherent states, called the von Neumann lattice, is complete and exists in the phase space with a unit cell of area $h$, the Planck constant. The uncertainty relation is naturally included in the lattice structure. This lattice then serves as a basis in quantum mechanics and has many applications[17,18]. The von Neumann procedure can also be used to construct a complete set of coherent states in the LLL[19]. These periodical coherent states (or von Neumann lattices), abbreviated by vN$_q$ lattices, exist in the coordinate space with a unit cell of area $q\phi/2\Omega$, where $q = 1,\ 2,\ 3....,\ \phi\ (\phi = h/M)$ is the flux quantum of circulation and $M$ is the inertial mass of an atom. There are $q\phi$ fluxes through a unit cell. The discrete coherent state in the LLL is

$$\psi_{mn}(\mathbf{r}) = \frac{1}{\sqrt{2\pi\ell^2}}\exp\left[-\frac{(\mathbf{r}-\mathbf{R}_{mn})^2}{4\ell^2} - \frac{i}{2\ell^2}\mathbf{r}\times\mathbf{R}_{mn}\bullet\hat{\mathbf{z}}\right],\qquad(1)$$

where $\mathbf{r} = (x,\ y)$ is the coordinate of a 2D plane, $m$ and $n$ are integers, $\ell = \sqrt{\phi/4\pi\Omega}$ is the magnetic length. Here, $\mathbf{R}_{mn} = m\mathbf{a}_1 + n\mathbf{a}_2$ defines a set of points of a Bravais lattice with primitive vectors $\mathbf{a}_1 = a(1,\ 0)$ and $\mathbf{a}_2 = a(u,\ v)$, where $a$ is the lattice constant, $u$ and $v$ are the geometric parameters of a lattice. $(u,\ v) = (1/2,\ \sqrt{3}/2)$ and $(0,\ 1)$, for triangular and square lattice, respectively.

Owing to $q\phi$ fluxes through a unit cell, the area, $va^2$, of a unit cell is quantized and $va^2 = q\phi/2\Omega$. The lattice constant becomes larger as $q\phi$ fluxes through a unit cell are increasing. The filling factor, $f$, of the system is given by $f = N_c / q$, where $N_c$ is the number of particles per unit cell. If $N_c < q$, the system is in the fractional quantum Hall regime and the quantum-Hall liquid can be more stable than the vortex-lattice state[20]. In this letter, we considered the cases where $N_c > q$, i.e., $f > 1$. For $f > 1$, the vortex-lattice state is believed to be the most stable state and this stability can be verified by numerical simulations.

The coherent state in Eq. (1) represents a cluster of bosonic atoms residing at a point of a Bravais lattice. Let $\Psi_q(\mathbf{r})$ be the macroscopic wave function of a vortex lattice with $q\phi$ fluxes through a unit cell. Owing to the completeness of coherent states $\psi_{mn}(\mathbf{r})$, we can express $\Psi_q(\mathbf{r})$ in terms of the von Neumann lattice and $\Psi_q(\mathbf{r}) = C_q \sum_{m,n} (-1)^{mnq} \psi_{mn}(\mathbf{r})$, where $C_q$ is a normalized constant given by the number of particles, $N$, in the system. The wavefunction $\Psi_1(\mathbf{r})$ for $q = 1$ was used to study the properties of vortex lattices of single vortices in rapidly rotating ultracold atomic gas. Here we have generalized the wavefunction to vortex lattices with multiple $\phi$ fluxes through a unit cell. Note that $\Psi_q(\mathbf{r} + \mathbf{R}) = (-1)^{jkq} e^{-i\mathbf{r} \times \mathbf{R} \cdot \hat{\mathbf{z}} / 2\ell^2} \Psi_q(\mathbf{r})$, where $\mathbf{R} = j\mathbf{a}_1 + k\mathbf{a}_2$, $j$ and $k$ are integers. Thus, $\Psi_q(\mathbf{r})$ is a quasiperiodic function[21] characterized by the vortex-lattice structure. It is the density distribution $\rho_q(\mathbf{r})$, defined by $\rho_q(\mathbf{r}) = \left| \Psi_q(\mathbf{r}) \right|^2$, that is a periodic function of the lattice, i.e., $\rho_q(\mathbf{r} + \mathbf{R}) = \rho_q(\mathbf{r})$.

A vortex-lattice state can be expanded by the Abrikosov or vN$_q$ lattice where the positions of vortices or clusters of bosonic atoms are considered as the basis of a vortex lattice, respectively. It seems that the vN$_q$ lattice is useful for the crystal formed by clusters of bosonic atoms and the vortex lattice is best described by the Abrikosov lattice. In fact, vN$_1$ lattices, one of vN$_q$ lattices with a $\phi$ flux through a unit cell, have the same physical properties as Abrikosov lattices (see Supplementary Fig. S1). Vortices are closely packed into a lattice structure. Atoms are distributed uniformly except near vortex sites. We can say that the Abrikosov and vN$_1$ lattice describe the same vortex lattice with a single vortex in a unit cell under the Landau and symmetric gauges, respectively.

To further show that vN$_1$ and Abrikosov lattices possess the same physical properties, we have to consider the nonlinear effect of contact interactions between particles for vN$_1$ lattices. A $\beta_A$ parameter, suggested by Abrikosov, will determine the strength of the nonlinear effect and relative stability of various vortex structures[2]. For vN$_q$ lattices, we can take this parameter as $\beta_q$, which depends on the $q\phi$ fluxes through a unit cell and is defined by $\beta_q = \left\langle \left| \Psi_q \right|^4 \right\rangle \bigg/ \left\langle \left| \Psi_q \right|^2 \right\rangle^2$, where $\langle F \rangle$ is the average of a function $F$ in a unit cell.

The $\beta_q$ value is larger as $\Psi_q(\mathbf{r})$ becomes more localized and peaked up[22]. $\beta_q$ values for various vN$_q$ lattices can be calculated and compare with the already known $\beta_A$ parameters of Abrikosov lattices. We find that $\beta_1^T = 1.1596$ and $\beta_1^S = 1.1803$ for triangular and square vN$_1$ lattices, respectively. These values are equal to $\beta_A$ values founded in type-II superconductors[22] and Bose-Einstein condensates[18,23]. In addition, we can find the ground state energies of vN$_q$ lattices with $q > 1$ (see Supplementary Table 1). Here we only show $\beta_q^T$ values of these lattices with a triangular structure: $\beta_2^T = 1.3390$, $\beta_3^T = 1.6015$ and $\beta_4^T = 2.0355$. These $\beta_q^T$ values are increasing with $q$ and higher than the $\beta_1^T$ value. We conclude that no $vN_q$ lattices with $q > 1$ and only stable Abrikosov lattices exist in various physical systems with contact interactions between particles[3-10]. Although this stability is consistent with various experiments, it is still an open question whether Abrikosov lattices are stable in systems with non-local interactions between particles[11,12].

Bose-Einstein condensates with non-local interactions, such as dipolar interactions[24], dipole-blockaded interactions[25-27] or Rydberg-dressed interactions[12,28], have been studied recently. The advances in manipulating atoms will reveal many interesting physical phenomena for systems with non-local interactions between atoms. Supersolid droplet crystals were discovered in a dipole-blockaded gas[25-27]. The dipole blockade was used to control the generation of collective excitations and manipulate the entanglement of these excitations for quantum information processing[29,30]. There exist bubble and supersolid-vortex crystals in some physical regimes of rotating dipolar[11] and Rydberg-dressed Bose gases[12], respectively. The existence of these crystal phases indicates that Abrikosov lattices in some physical regimes may not be stable vortex lattices in systems with non-local interactions between particles. No theory explains their existence and studies their physical properties, except numerical simulations. We will see that a bubble crystal or a supersolid-vortex crystal is just one of kinds of $vN_q$ lattices.

Until know there is no solid evidence and only speculation that $vN_q$ lattices may exist in the systems with non-local interactions between particles. To see whether $vN_q$ lattices with $q > 1$ exist in any physical systems, we consider the physical system of rapidly rotating dipole-blockaded (RRDB) gases as the candidate to realize these lattices. We assume that this gas is a condensate in a pancake shaped geometry and the interaction energy per particle is smaller than the trap energy. The z-direction degree of freedom of the condensate is then frozen into the ground state of the harmonic oscillator in the z-direction. Due to the centrifugal force of rapidly rotating the system, the density of RRDB gases is spreading and reduced. The system is then approaching the weak-interaction limit and falling into the LLL regime[8]. The ground-state energies of vN$_q$ lattices in RRDB gases can be easily calculated via $\Psi_q(\mathbf{r})$.

For a fixed rotating frequency and dipole-blockade radius[25-27], $b$, we compare ground-state energies of vN$_q$ lattices for different geometrical structures and $q$ values to find the most stable lattice. We find that the Abrikosov or vN$_1$ lattices are stable if $b/\ell \leq 1.82$ and these lattices are: a triangular Abrikosov lattice $(0 \leq b/\ell \leq 1.46)$; a square Abrikosov lattice $(1.46 \leq b/\ell \leq 1.59)$; a stripe phase $(1.59 \leq b/\ell \leq 1.82)$. The phase diagram of stable vortex lattices of RRDB gases is shown in Fig. 1. The $q$ number of labeling the lowest-energy state of these lattices is increasing step by step as $b$ becomes larger. In the regime of $b/\ell \geq 1.82$, We find a sequence of vN$_q$ lattices with $q > 1$: triangular vN$_2$ lattices $(1.82 \leq b/\ell \leq 2.0)$; square vN$_2$ lattices $(2.0 \leq b/\ell \leq 2.2)$. As $b/\ell \geq 2.2$, the triangular vN$_q$ lattices are the stable states in the regimes given by: vN$_3$ lattices $(2.2 \leq b/\ell \leq 2.5)$, vN$_4$ lattices $(2.5 \leq b/\ell \leq 2.9)$, vN$_5$ lattices $(2.9 \leq b/\ell \leq 3.2)$, vN$_6$ lattices $(3.2 \leq b/\ell \leq 3.5)$ and vN$_7$ lattices $(3.5 \leq b/\ell \leq 3.8)$.

The vN$_q$ lattices with $q > 1$ are energetically favorable in the regime of larger dipole-blockade radiuses. Their density and vortex distributions are shown in Fig. 2. To see the vortex distribution clearly, we drew the log-density distributions, $\mathrm{Log}_e[\rho_q(\mathbf{r})]$, of triangular vN$_q$ lattices with $q = 2$ to 5. The blue dashed lines outline the area of a unit cell. Atoms of vN$_q$ lattices with $q > 1$ are clustered periodically and surrounded by vortices whose number is increasing with $b$. Unlike vN$_1$ lattices, the quantum pressure from the centrifugal forces of vortices surrounding a vN$_q$–lattice site becomes larger and strong enough to force atoms being clustered periodically as the number of these vortices is increased. These lattices then look like bubble crystals[11] and supersolids[12] found in rapidly rotating dipolar and Rydberg-dressed gases, respectively.

There is a general trend that the number of single vortices in a unit cell is equal to the $q$ number of labelling vN$_q$ lattices. This trend is invalid for the triangular vN$_4$ lattice due to its geometrical structure. While there exist 4 single vortices in a unit cell of the square vN$_4$ lattice, the triangular vN$_4$ lattice has 2 vortices with double fluxes in a unit cell (see Supplementary Fig. S2). These vortex distributions exhibit the same pattern as the triangular bubbles[11], which have multiple fluxes associated with each bubble, in rapidly rotating dipolar gases. In fact, bubble crystals in rapidly rotating dipolar gases are triangular vN$_q$ lattices with $q > 1$. From vN$_1$ lattices being Abrikosov lattices, we conclude that Abrikosov lattices, bubble and supersolid-vortex crystals are sub-lattices of vN$_q$ lattices, which are generalized vortex lattices.

Indeed, the calculations of ground-state energies of vN$_q$ lattices have shown that there is a series of vN$_q$ lattices existing in the physical system of RRDB gases. But, we need extra evidences to confirm their existence. We have done numerical simulations to check the correctness of the phase diagram of triangular vN$_q$ lattices for several $b$ values. For a fixed $b$ value and $\Omega/\omega = 0.9$, where $\omega$ is the radial trapping frequency, the stable state of RRDB gases is obtained by solving the Gross-Pitaeviskii equation numerically. The density distributions from numerical simulations are plotted in log scale and shown in Fig. 3. These distributions exhibit similar patterns as vN$_q$ lattices.

Atoms in the stable vortex lattice are clustered periodically and surrounded by vortices whose number is increasing with $b$. Clustered atoms arranged in a triangular lattice. Owing to the finite-size effect from the trapping potential, the average lattice constants of vortex lattices from simulations are a little bit smaller than the lattice constants of corresponding infinite $vN_q$ lattices.

From the vortex distribution in a unit cell outlined by blue dashed lines in Fig. 3, we can count how many $q\phi$ fluxes through a unit cell. Vortices in a unit cell is increasing step by step with $b$. The numerical $q$ number of fluxes through a unit cell for a fixed $b$ and $\Omega/\omega = 0.9$ is marked by a red square in Fig. 1. This number is consistent with the $q$ number of labeling the lowest-energy state of triangular $vN_q$ lattices in Fig. 1. From these density distributions, we find that triangular vortex lattices with multiple fluxes through a unit cell are energetically favorable in the regime of larger dipole-blockade radiuses. The stable vortex lattices found by numerical simulations are consistent with theoretical calculations of $vN_q$ lattices. The number of flux quantums through a unit cell of the steady state found by numerical simulations is also increasing with $b$. These vortex lattices are the representations of $vN_q$ lattices. The von Neumann lattices as generalized vortex lattices can be physically realized by vortex lattices in RRDB gases.

## METHODS SUMMARY:

Using the Gross-Pitaevskii (GP) approach for trapped dipole-blockaded Bose gases, the GP equation can be written as:

$$i\hbar \frac{\partial}{\partial t} \Psi(\mathbf{r}, t) = \left[ \frac{1}{2M} \left( \mathbf{p} - M\,\Omega\,\hat{\mathbf{z}} \times \mathbf{r} \right)^2 + \frac{1}{2} M \left( \omega^2 - \Omega^2 \right) r^2 \right] \Psi(\mathbf{r}, t)$$
$$+ \int d^2 \mathbf{r}' V(\mathbf{r} - \mathbf{r}') \left| \Psi(\mathbf{r}') \right|^2 \Psi(\mathbf{r}, t), \tag{2}$$

where $\omega$ and $\Omega$ are the trap and rotation frequencies, respectively. $V(r)$ in Eq. (2) is the dipole-blockaded potential between particles:

$$V(r) = \begin{cases} \dfrac{D}{b^3} & r \le b \\[2mm] \dfrac{D}{r^3} & r > b \end{cases}. \qquad (3)$$

Here, $D$ is the dipole interaction strength and $b$ is the blockaded radius. We consider a variational macroscopic wave function of the form $\Psi_q(\mathbf{r}) = C_q \sum_{m,n} (-1)^{mnq} \psi_{mn}(\mathbf{r})$, where $C_q$ is a normalized constant given by the number of particles, $N$, in the system. Since the variational wave function is a superposition of coherent states in the lowest Landau level, the kinetic plus trap potential energy per particle for the state $\Psi_q(\mathbf{r})$ is a constant, $\hbar\Omega$, in the $\Omega \to \omega$ limit. Therefore, the ground state of vN$_q$ lattices for a fixed blockade radius is determined by the GP interaction energy $E_{int}$, where

$$E_{int} = \frac{1}{2} \int d^2\mathbf{r}\, d^2\mathbf{r}'\, V(\mathbf{r}-\mathbf{r}') \left|\Psi_q(\mathbf{r}')\right|^2 \left|\Psi_q(\mathbf{r})\right|^2. \qquad (4)$$

By comparing $E_{int}$ for lattices with different geometric structures and $q$ fluxes through a unit cell, we then find the phase diagram of vN$_q$ lattices for different blockade radiuses.

To verify the phase diagram of Fig. 1, we solve Eq. (2) numerically. The state at a late time is obtained by applying the split operator method on the time evolution of a wave function. For a fixed rotating frequency, $\Omega/\omega = 0.9$ and interaction strength, $ND = 1500$, we chose one of vN$_q$ lattices shown in Fig. 2 as the initial state and let the state evolve in imaginary time by substituting $t$ into $-it$. The system then approaches to a stable state gradually during time evolving.

**Supplementary Information** is available in the online version of the paper.

**Acknowledgements** This work was supported by Ministry of Science and Technology of Republic of China (NSC102-2112-M-034-001-MY3). S.C.C. thanks the support of the National Center for Theoretical Sciences of Taiwan during visiting the center.

**Author Contributions** S.C.C. developed the theory, carried out the theoretical analysis and wrote the paper. S.D.J. carried out the numerical simulations and analyzed the numerical results.

**Author Information** Reprints and permissions information is available at www.nature.com/reprints. The authors declare no competing financial interests. Readers are welcome to comment on the online version of the paper. Correspondence and requests for materials should be addressed to S.C.C. (sccheng@faculty.pccu.edu.tw).

# Figure captions

**Figure 1: Phase diagram of triangular vN$_q$ lattices**. The horizontal and perpendicular axes are the blockade radius and the number of fluxes through a unit cell of vN$_q$ lattices, respectively. Here, **T**, **S**, and **R** denotes triangular, square and stripe lattices, respectively. The solid blue lines and red squares are the results from ground-state calculations and numerical simulations of rapidly rotating dipole-blockaded gases, respectively. The particle log-density distributions of stable vortex lattices corresponding to blockade radiuses marked by red squares are shown in Fig. 3.

**Figure 2: Contour plots of log-scaled particle density distributions of triangular vN$_q$ lattices**. The low-density regime shown by the black area indicates the location of a vortex and the bright area is the high-density regime. The blue dashed lines outline the area of a unit cell. The log-density distributions of triangular vN$_2$ and vN$_3$ lattices are shown in (a) and (b) diagrams, respectively. The (c) and (d) diagrams show the log-density distributions of triangular vN$_4$ and vN$_5$ lattices, separately. The lattice constants $a/\ell = 3.81$, 4.67, 5.39 and 6.02 for triangular vN$_2$, vN$_3$, vN$_4$ and vN$_5$ lattices, respectively.

**Figure 3: Contour plots of log-scaled particle density distributions of stable vortex lattices**. The low-density regime shown by the black area indicates the location of a vortex and the bright area is the high-density regime. The blue dashed lines outline the area of a unit cell. The density distributions of stable vortex lattices for blockade radiuses $b/\ell = 1.88$, 2.28, 2.68 and 3.02 are shown in diagrams (a), (b), (c) and (d), respectively. The average lattice constants $a/\ell = 3.83$, 4.48, 5.18 and 5.80 for (a), (b), (c) and (d) diagrams, respectively.

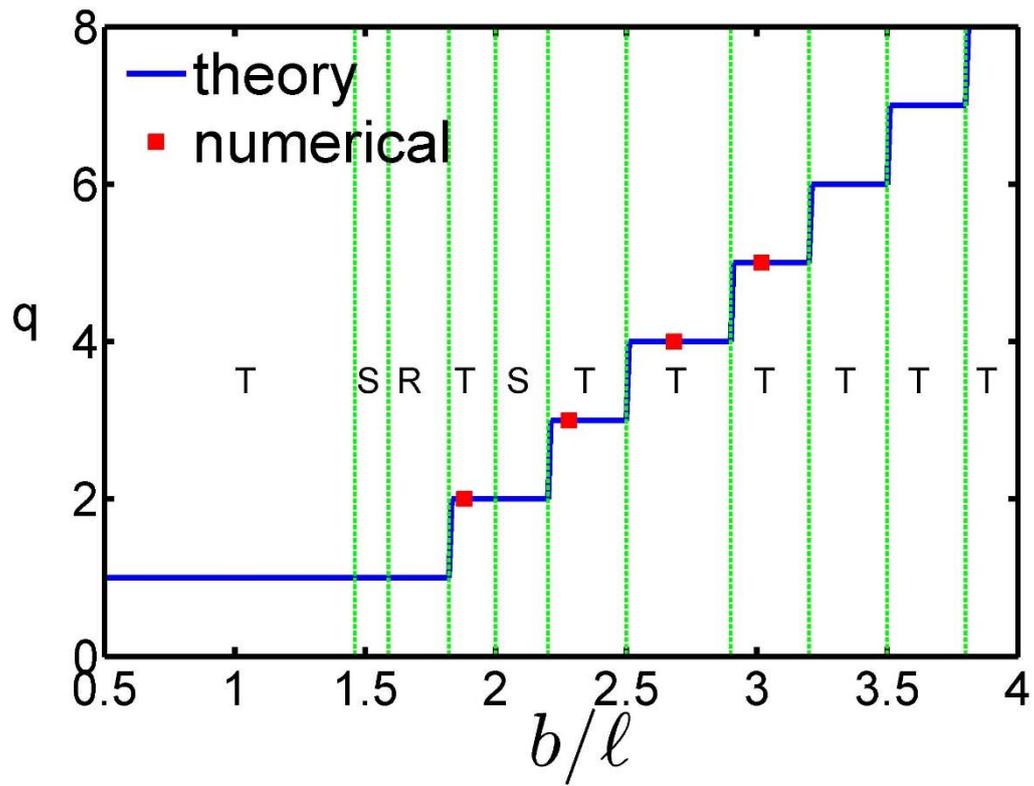

Figure 1

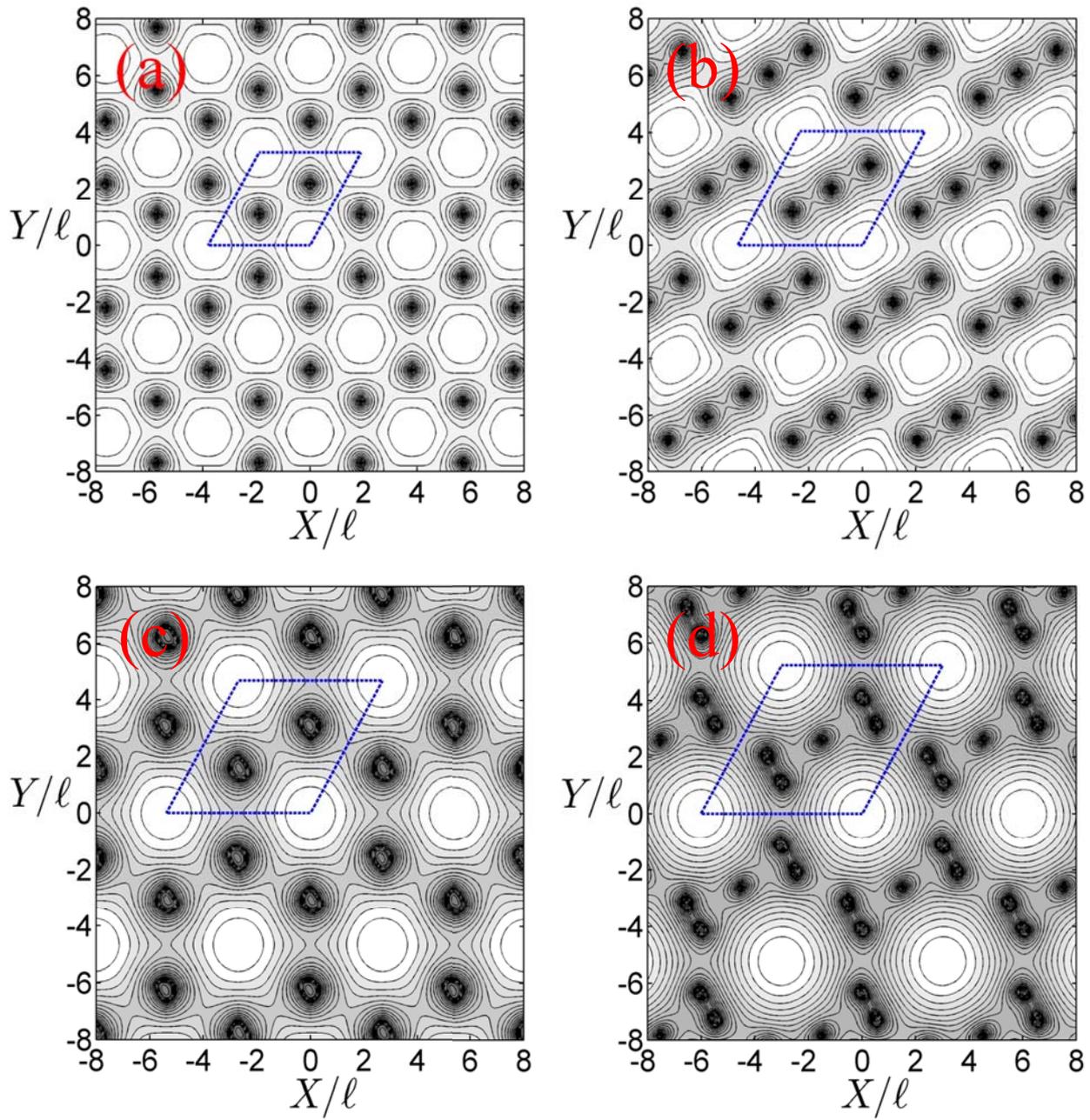

Figure 2

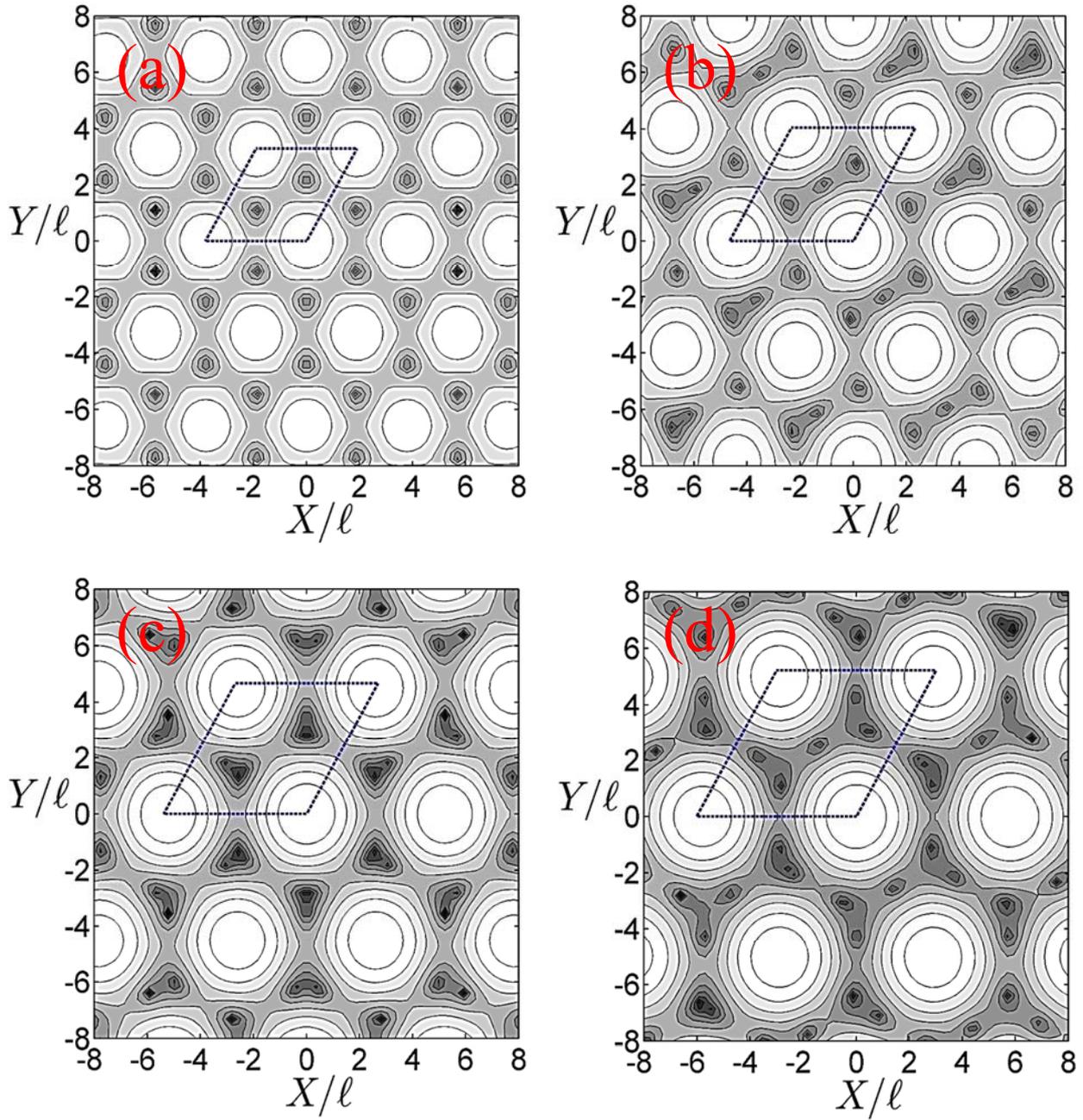

Figure 3

# Supplementary Information for "Physical Realization of von Neumann Lattices in Rotating Dipole-blockaded Bose Gases"

Szu-Cheng Cheng[1] & Shih-Da Jheng[2]

*[1]Dept. of Optoelectric Physics, Chinese Culture University, Taipei 11114, Taiwan, ROC*

*[2]Dept. of Photonics and Institute of Electro-Optical Engineering, National Chiao Tung University, Hsinchu 30010 ,Taiwan, ROC*

Using $vN_q$ lattices, we can easily describe vortex lattices with $q\phi$ fluxes through a unit cell, where $q > 1$. $vN_q$ lattices are generalized vortex lattices which become Abrikosov lattices as $q = 1$. We show the density distributions, $\rho_1(\mathbf{r})$, of $vN_1$ lattices in Fig. S1. These distributions have the same characters owned by Abrikosov lattices[2]. Because of $vN_1$ lattices having only a single vortex in a unit cell, the quantum pressure from the centrifugal forces of vortices around a $vN_1$−lattice site is not strong enough to force atoms being clustered. Atoms are then distributed uniformly except near vortex sites. Therefore, the $vN_q$ lattices become Abrikosov lattices as $q = 1$. This fact is not surprising because the physical phenomena are independent on the choice of a gauge. $vN_1$ lattices are the representations of Abrikosov lattices in the symmetric gauge.

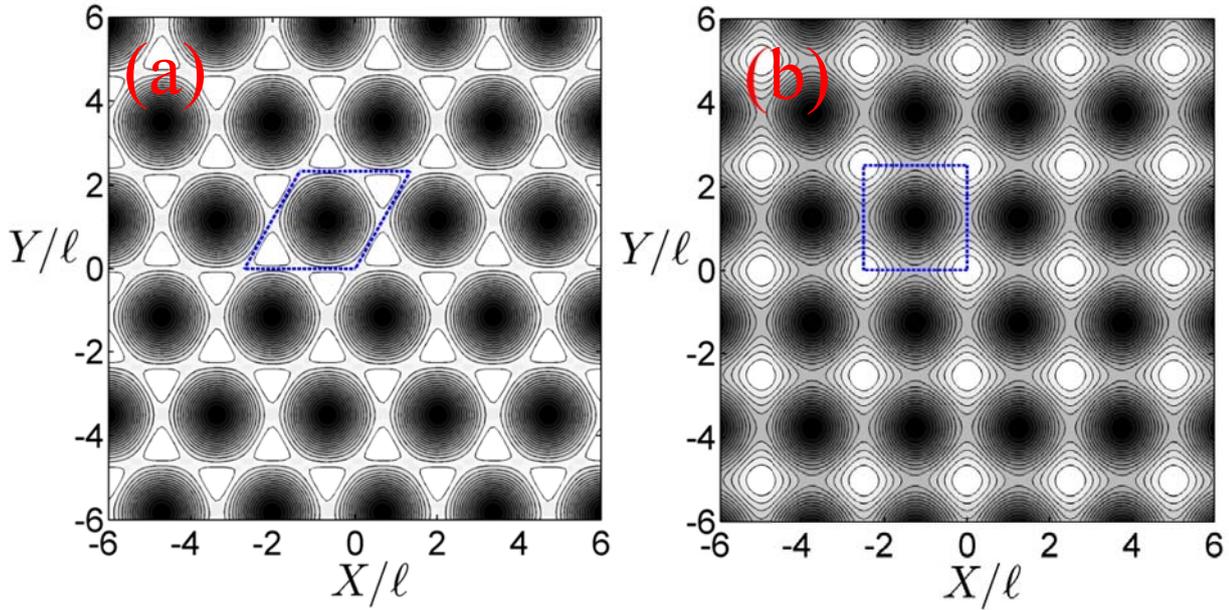

**Figure S1: Contour plot of particle density distributions of vN$_1$ lattices.** vN$_1$ lattices are triangular and square structures in (a) and (b) diagrams, respectively. The low-density regime shown by the black area indicates the location of a vortex and the bright area is the high-density regime. The blue dashed lines outline the area of a unit cell. There is a flux quantum through a unit cell in (a) and (b) diagrams. The lattice constants $a/\ell = 2.69$ and $2.51$ for (a) and (b) diagrams, respectively.

For systems with contact interactions between particles, the ground-state energy of a vortex lattice is proportional to the Abrikosov $\beta$ parameter. In Table S1 we list several Abrikosov $\beta^T$ and $\beta^S$ parameters for triangular and square structures of vN$_q$ lattices, respectively. We find that the $\beta^T$ parameter of a triangular vN$_q$ lattice is smaller than the $\beta^S$ parameter of a square vN$_q$ lattice when $q\phi$ fluxes through a unit cell are fixed.

Therefore, the triangular $vN_q$ lattice has a lower energy and more stable than the square $vN_q$ lattice. Abrikosov $\beta$ parameters are also increasing with the number of fluxes through a unit cell being increased. The triangular $vN_q$ lattice with $q = 1$ shows the smallest $\beta$ parameter, i.e., $\beta^T = 1.1596$. The triangular Abrikosov lattice or $vN_1$ lattice is then more stable than any other vortex lattices in physical systems with contact interactions between particles.

**TABLE S1.** Abrikosov $\beta^T$ and $\beta^S$ parameters for triangular and square structures of von Neumann lattices, respectively. The symbol $q$ is the number of fluxes through a unit cell.

| $q$ | $\beta^T$ | $\beta^S$ |
|-----|-----------|-----------|
| 1 | 1.1596 | 1.1803 |
| 2 | 1.3390 | 1.4248 |
| 3 | 1.6015 | 1.7015 |
| 4 | 2.0355 | 2.0598 |
| 5 | 2.5046 | 2.5155 |

Although the square $vN_q$ lattice is stable in the regime $2.0 \leq b/\ell \leq 2.2$, we show density distributions of square $vN_q$ lattices for $q = 2$ to 5 in Fig. S2. The geometric structure of a $vN_q$ lattice strongly affects the vortex distribution in a unit cell. While there exist 2 and 3 single vortices in a unit cell of triangular $vN_2$ and $vN_3$ lattices, respectively, the square $vN_2$ ($vN_3$) lattice has a vortex with double (triple) fluxes in a unit cell. For square $vN_4$ and $vN_5$ lattices, there are 4 and 5 single vortices in a unit cell, respectively.

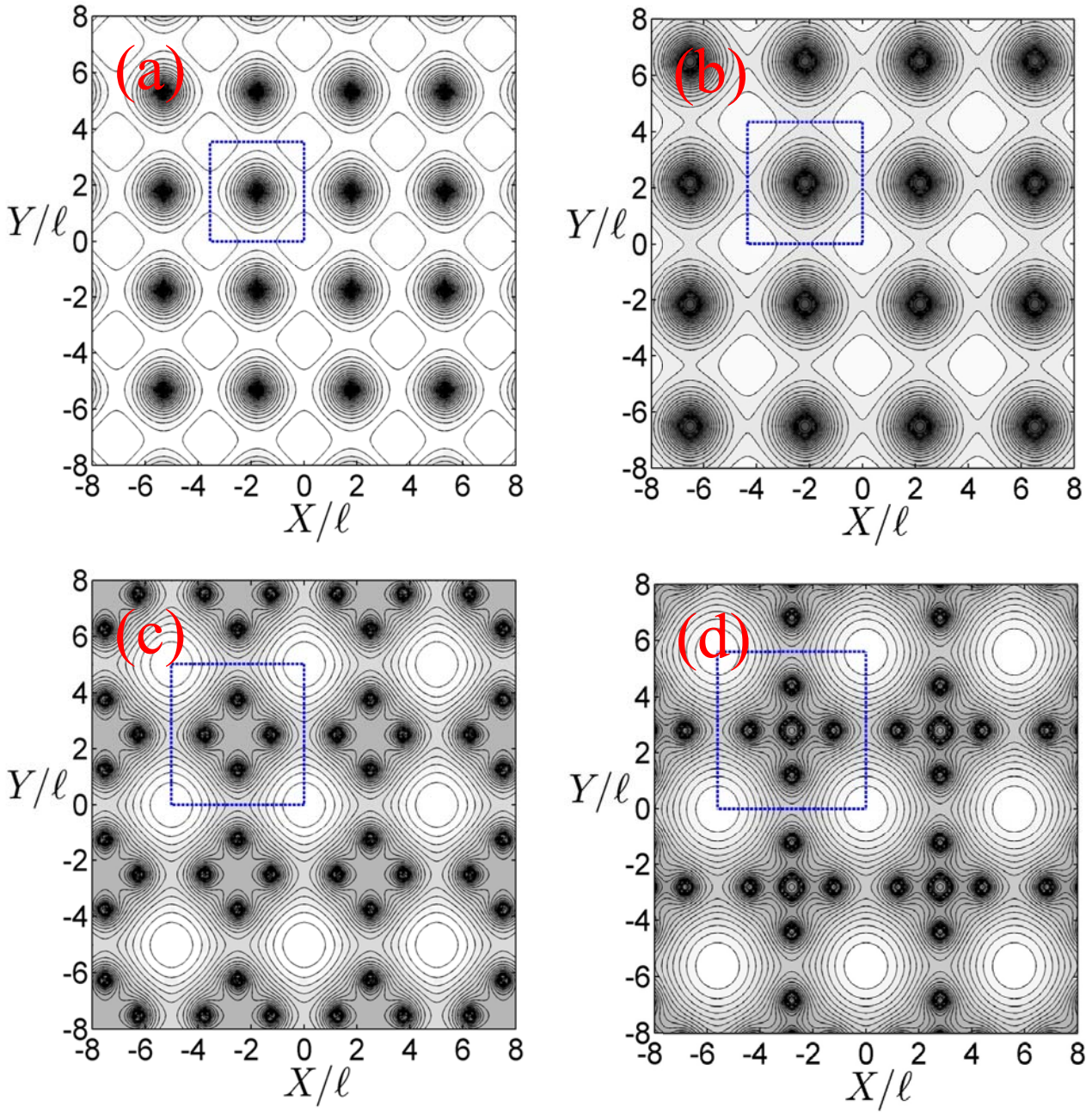

**Figure S2: Contour plots of log-scaled particle density distributions of square $vN_q$ lattices**. The low-density regime shown by the black area indicates the location of a vortex and the bright area is the high-density regime. The blue dashed lines outline the area of a unit cell. The log-density distributions of square $vN_2$ and $vN_3$ lattices are shown in (a) and (b) diagrams, respectively. The (c) and (d) diagrams show the log-density distributions of square $vN_4$ and $vN_5$ lattices, separately. The lattice constants $a/\ell = 3.54$, 4.34, 5.01 and 5.60 for square $vN_2$, $vN_3$, $vN_4$ and $vN_5$ lattices, respectively.